\documentstyle[prl,aps,multicol]{revtex}

\begin{document}

\title{Quantum algorithms which accept hot qubit inputs}

\author{Xinlan Zhou{$^{1,2}$}, Debbie W. Leung{$^{3,2}$}, 
	Isaac L. Chuang{$^2$}}

\address{\vspace*{1.2ex}
        {$^1$ Applied Physics Department, Stanford University, 
                  Stanford, CA 94305}\\[1.2ex]
        {$^2$ IBM Almaden Research Center,  San Jose, CA 95120}\\[1.2ex]
	{$^3$ Physics Department, Stanford University, 
		  Stanford, CA 94305}\\[1.2ex]
}
                                     
\date{\today \vspace*{-.3cm}}

\def\<{\langle}
\def\>{\rangle}
\def\be{\begin{equation}}
\def\ee{\end{equation}}
\def\bea{\begin{eqnarray}}
\def\eea{\end{eqnarray}}
\def\lbL{ \left[\rule{0pt}{2.4ex} }
\def\rbL{ \right] }

\newcommand{\ket}[1]{\mbox{$|#1\rangle$}}
\newcommand{\bra}[1]{\mbox{$\langle #1|$}}
\newcommand{\mypsfig}[2]{\psfig{file=#1,#2}}
\maketitle

\begin{abstract}
Realistic physical implementations of quantum computers can entail
tradeoffs which depart from the ideal model of quantum computation.
Although these tradeoffs have allowed successful demonstration of
certain quantum algorithms, a crucial question is whether they
fundamentally limit the computational capacity of such machines.  We
study the limitations of a quantum computation model in which only
ensemble averages of measurement observables are accessible.
Furthermore, we stipulate that input qubits may only be prepared in
highly random, ``hot'' mixed states.  In general, these limitations
are believed to dramatically detract from the computational power of
the system.  However, we construct a class of algorithms for this
limited model, which, surprisingly, are polynomially equivalent to
the ideal case.  This class includes the well known Deutsch-Jozsa
algorithm.
\end{abstract}

\begin{multicols}{2}

The discovery of fast quantum algorithms
\cite{DeutschJozsa,Shor94,Simon94,Grover} with no classical counterparts has
led to an increasing interest in the experimental realization of quantum
computers.  However, in practice, it is expected that realistic
implementations of these machines will suffer from imperfections which depart
from the ideal model of quantum computation.  Ideally, standard quantum
computation starts with a system prepared in a pure state such as $|00\cdots
0\>$, and unitary transforms are applied systematically as prescribed by the
algorithm, obtaining some final state $|\psi\>$.  When measured, this
state collapses with high probability into a small number of possible
outcomes, which reveal the desired answer.

In contrast, recent demonstrations of simple quantum
algorithms\cite{Chuang98a,Chuang98b,Jones98a,Jones98b,Jones98c} have
employed a far different model of quantum computation, in which the
system is a bulk ensemble of a large (but finite) number of quantum
computers operating in parallel, prepared initially in a highly mixed
state, and with the only accessible measurement result being an
average of the observable over the ensemble.  Despite these stringent
limitations, it has been shown how quantum computation can indeed be
performed, by creating ``effective pure states''
\cite{Gershenfeld97,Cory97x,Chuang97e,Knill97}, and by modifying
quantum algorithms to produce deterministic results which do not
average away~\cite{Gershenfeld97}.  However, it is understood that
using present techniques, this ability comes at a cost: either an
exponential reduction in signal strength as a function of the size of
the computer, or a linear reduction in the number of usable
qubits~\cite{Schulman98}, with all other resources held constant.

These observations lead to two important points motivating the present work.
First, it is an open question whether or not the limited ``bulk quantum
computer'' (BQC) model described above is as powerful as the standard quantum
computer (SQC) model.  Previous work has shown that BQC with thermal inputs is
strictly less powerful than SQC in the presence of oracles \cite{Knill99}.
Nontheless, it is still possible that new techniques will be found which make
them equivalent; this has not been proven impossible, but such a result would
be unexpected.  Second, the limitations of the BQC model are not fundamental
-- they arise from practical considerations which significantly simplify
experimental realization.  Might there be other kinds of restrictions to the
SQC model which apply to the most important quantum algorithms, and also
simplify implementation while providing a provably polynomially equivalent
model of quantum computation?

In this Letter, we prove the existence of non-trivial overlap between
the BQC (with thermal inputs) and SQC models: a surprising subclass of
quantum algorithms for which BQC and SQC are polynomially equivalent.
This includes the well known Deutsch-Jozsa (DJ) algorithm.  The
equivalence arises from the robustness of the algorithms to
independent bit flip errors which occur before the computation.  Our
result does not resolve the general question of equivalence between
BQC and SQC, but it provides new insight into the degrees of freedom
which quantum algorithms allow different computational models.

It is convenient to begin with the Deutsch-Jozsa algoithm\cite{DeutschJozsa}
(as improved in \cite{Cleve97}), which solves the following
problem.
Alice has a function $f:\{0,1\}^n \rightarrow \{0,1\}$ which is unknown to
Bob. She promises that $f$ is either constant or balanced ($f(x) = 0$ for
exactly half of the inputs and $f(x) = 1$ otherwise).  Bob wants to find out
whether $f$ is constant or balanced with the minimum number of queries.
With a deterministic classical computer, $2^{n-1}+1$ queries are required in
the worst case to determine the type of the function.  However, in the SQC
model, the following algorithm can be performed:
1. Starting from the state $|0\>^{\otimes n}|1\>$, Bob prepares a
query input register in a superposition of all basis states using the
Hadamard transformation $H$, and an ancilla qubit in the state $H|1\>
= (\ket{0}-\ket{1})/ \sqrt{2}$.  Bob sends both the input and ancilla
to Alice.
2. Alice evaluates $f(x)$, stores the answer in the ancilla
qubit and sends the qubits back to Bob afterwards.
3. Bob applies $H$ to the input register and the ancilla qubit.
These three steps can be represented by the following expressions:
\bea
	\ket{0^n}\ket{1} 
	&\stackrel{1. H^{\otimes n}\otimes H} {\longrightarrow} &
		\frac{1}{\sqrt{2^{n+1}}} 
		\sum_{x} 
		\ket{x}(\ket{0}-\ket{1})
\label{eq:step1}
\\	&\stackrel{2. U_f} {\longrightarrow} & 
		\frac{1}{\sqrt{2^{n+1}}} 
		\sum_{x} 
		(-1)^{f(x)} \ket{x} (\ket{0}-\ket{1})
\label{eq:step2}
\\	&\stackrel{3. H^{\otimes n}\otimes H} {\longrightarrow} &  
		\frac{1}{2^n} 
		\sum_{x}
		\sum_{y}
		(-1)^{x\cdot y \oplus f(x)} \ket{y}\ket{1}
\label{eq:step3}
\,,
\eea
where all summations are over $\{0,1\}^{n}$. 
The final state of the input register is therefore 
\be
	\ket{\phi_f^{SQC}} = \frac{1}{2^n} \sum_y 
		\sum_x (-1)^{f(x) \oplus
		x\cdot y} \ket{y} = \sum_y g(y) \ket{y}
\label{eq:firstfinal}
\,,
\ee
with 
\be
	g(y) = \frac{1}{2^n}\sum_{x} (-1)^{f(x) \oplus x\cdot y}
\,.
\ee
If $f$ is constant, $|g(y)|^2 = \delta(y,0)$; 
if $f$ is balanced, $|g(0)|^2 = 0$.
When Bob projects the qubits along the basis states, he obtains a
definite value of $y$ to decide whether $f$ is constant or balanced.
The problem is therefore solved with only one query in the SQC model.

Does the DJ algorithm also work efficiently in the BQC model?  First,
let us consider a less restrictive model than BQC, which we shall term
BQC$_P$.  This model is the same as BQC, but with the initial state of
the quantum computers prepared in the pure state $|00\cdots0\>$.  For
this model, a resolution to our question comes from looking at the DJ
algorithm from the following, different point of view.

Projection onto the basis states, which are eigenstates of $\sigma_z$ 
($\sigma_z \ket{x} = (2x-1) \ket{x}, x\in \{0,1\}$), performs mesaurements 
of the eigenvalues $\pm 1$ from each qubit. 
These eigenvalues are $-1$ for all qubits when $f$ is constant and $+1$ for at
least one qubit when $f$ is balanced.
Thus the DJ algorithm can be viewed as distingushing these two possible 
outcomes.  
For BQC$_P$, the output from the $i^{th}$ qubit is the expectation
value of $\sigma_{z_i}$, given by
\be
        E^{BQC_P}_i \equiv \<\phi_f^{BQC_P}|\sigma_{z_i}\ket{\phi_f^{BQC_P}} = 
                \sum_{y} (2 y_i-1) |g(y)|^2
\,, 	
\label{eq:ebqcpi}
\ee
where $y_i$ is the $i^{th}$ bit of $y$. 
If $f(x)$ is constant, $E^{BQC_P}_i = -1$ for all $i$.  
If $f$ is balanced, $\exists i$ such that $E^{BQC_P}_i > -1$. 
Due to the ensemble average over measurement results required in BQC$_P$, the
two types of functions may not be distinguishable, as they are in SQC.
The principle question is, can the two cases be distinguished without
exponentially increasing the space-time complexity?

We answer in the affirmative by analyzing the following measure of
distinguishibility, defined as
\be
       \epsilon(n)= \min_{f(x)_{bal}} \left[
		    \max_i \left[ 
	            E^{BQC_P}_{i,bal} - E^{BQC_P}_{i,const}\right]  
                \right] 
\label{eq:ep}
\,.
\ee
That is, $\epsilon(n)$ is the difference in the signals from the
constant function and the {\em worst} balanced function Alice can apply.
We show that BQC$_P$ and SQC are polynomially equivalent by proving the
following fact that $\epsilon(n)$ is lower bounded by the
inverse of a polynomial in $n$. \\ 
{\em Fact~1}: $\forall n, \epsilon(n) \ge 2/n$. \\
{\em Proof}:   
Since $E^{BQC_P}_{i,const} = -1$ is independent of $f$ and the maximum over 
$i$ is lower bounded by the average, from Eq.(\ref{eq:ep}), 
\be 
       \epsilon(n) \ge \min_{f(x)_{bal}} \left[ 
		\frac{1}{n} \sum_i \left[E^{BQC_P}_{i,bal} \right] + 1   
                \right] 
\label{eq:epb}
\,.
\ee
Using Eq.~(\ref{eq:ebqcpi}),
\bea
	    \sum_{i=1}^n E_{i,bal}^{BQC_P}
	&=& \sum_{i=1}^n \sum_{y=0}^{2^n-1}  (2y_i-1)|g(y)|^2
\label{eq:name1}
\\	&=& 2 \sum_{y=1}^{2^n-1}|g(y)|^2 (\sum_{i=1}^n y_i)  - n
\label{eq:name2}
\,,
\eea 
where the crucial fact $|g(0)|^2=0$ inherent in DJ is used to omit the
$y=0$ term in Eq.~(\ref{eq:name2}).  We have also used the
normalization $\sum_{y=0}^{2^n-1}|g(y)|^2 = 1$.
Since $\forall y \geq 1$, $\sum_{i=1}^n y_i \geq 1$, it follows that  
\bea
	\frac{1}{n} \sum_{i=1}^n E_{i,bal}^{BQC_P}	
	 \geq  \frac{2}{n} \sum_{y=0}^{2^n-1}|g(y)|^2 - 1 
	= \frac{2}{n} - 1 
\label{eq:ave}
\,. 
\eea
The desired result is obtained by substituting Eq.(\ref{eq:ave}) into
Eq.(\ref{eq:epb}). 
$\Box$ \\
Since $\epsilon(n) \ge 2/n$, it can be classically amplified to become
detectable in time $O(n^2)$.  Thus, BQC$_P$ can solve the DJ problem
with $O(n^2)$ queries, exponentially better than deterministic
classical computation.

Now we return to the BQC model, which is more realistic than BQC$_P$, as
thermal initial states are typically more experimentally feasible to prepare
than pure states.
The thermal initial state of BQC can be conveniently expressed in the
notation of~\cite{Chuangsw} as $[in\> = \oplus_k \sqrt{p_k} |k\>$,
where $[\cdot\>$ denotes a probabilistic mixture over some ensemble of
pure states, $\ket{k}=\ket{k_1,k_2,\ldots,k_n}$ with $k_i\in \{0,1\}$,
and $p_k$ is the probability of having the pure state $\ket{k}$.  When
the $i^{th}$ thermal qubit being $|0\>$ has probability $q_i$,  $p_k =
\Pi_{i} q_i^{1-k_i} (1-q_i)^{k_i}$ represents an uncorrelated initial
distribution.
Applying the DJ algorithm to the thermal state directly, one obtains 
\be
	[\phi_f^{BQC}\> = 
	\oplus_k \sqrt{p_k} H^{\otimes n} \tilde{U}_f H^{\otimes n}\ket{k}
\label{eq:finalthermal}
\,,
\ee
where $\tilde{U}_f$ denotes the restriction of $U_f$ to the input register.
For simplicity, we have assumed that the ancilla qubit is pure.  It will be 
clear why it does not affect our conclusion.  
We prove in the following that BQC and BQC$_P$ are polynomially equivalent 
for the DJ algorithm. \\
{\em Fact 2:} 
For the DJ algorithm, all BQC measurement results are proportional to those
from BQC$_P$ with constant proportionalities independent of $n$.
\\
{\em Proof:}
First, we note that for each qubit, $\ket{k_i} = \sigma_{x_i}^{k_i}
\ket{0}$, therefore $\ket{k} = \ket{k_1,\ldots,k_n} =
\sigma_{x_1}^{k_1} \ldots \sigma_{x_n}^{k_n}\ket{00 \cdots 0}$, which is
abbreviated as $\sigma_x^k\ket{0}$.
Second, from Eq.~(\ref{eq:step2}), it is clear that $\tilde{U}_f$ only
changes the relative signs of the states $|x\>$; hence,
$\tilde{U}_f$ commutes with $\sigma_z$ on any input qubit.
It follows that $[\sigma_x^k, H^{\otimes n} \tilde{U}_f H^{\otimes n}]$
= $H^{\otimes n} [\sigma_z^k, \tilde{U}_f] H^{\otimes n} = 0$. 
The output state of the DJ
algorithm in BQC as given by Eq.~(\ref{eq:finalthermal}) can therefore be
written as:
\bea
	[\phi_f^{BQC}\> 
	&=& \oplus_k \sqrt{p_k} 
		H^{\otimes n} \tilde{U}_f H^{\otimes n} \sigma_x^k \ket{0}
\\	&=& \oplus_k \sqrt{p_k} 
		\sigma_x^k H^{\otimes n}  \tilde{U}_f H^{\otimes n} \ket{0}
\\	&=& \oplus_k \sqrt{p_k} 
		\sigma_x^k \ket{\phi_f^{BQC_P}}
\,.
\eea
Hence, the signal from the $i^{th}$ qubit is given by \bea E^{BQC}_i
&=& \sum_k p_k \<\phi_f^{BQC_P}| \sigma_x^k \sigma_{z_i} \sigma_x^k
\ket{\phi_f^{BQC_P}} \\ &=& \sum_k p_k (-1)^{k_i} \<\phi_f^{BQC_P}|
\sigma_{z_i} \ket{\phi_f^{BQC_P}} \\ &=& \sum_k p_k (-1)^{k_i}
E^{BQC_P}_i \\ &=& ({\rm Pr}(k_i=0)- {\rm Pr}(k_i = 1)) E^{BQC_P}_i \\
&=& (2q_i -1) E^{BQC_P}_i \, \eea where we have used $q_i = {\rm
Pr}(k_i=0)$.  
$\Box$ \\
From Facts~1 and~2, it follows that $\epsilon(n)$ is at least
$(2q'-1)\frac{2}{n}$ for $q' = min(q_1,\ldots, q_n)$. 
The fact that the ancilla is a thermal state will only reduce
$\epsilon(n)$ by a constant factor of two.
This establishes that the DJ algorithm operating with the BQC model
also requires only $O(n^2)$ queries, produces an output with no loss
in signal strength, and achieves an exponential speed up relative
to deterministic classical computation.

This surprising result can be generalized to define an entire class of quantum
algorithms which are polynomially equivalent on both the BQC and SQC models.
We shall refer to these as ``hot qubit algorithms'' (HQA).
Let the initial state of the BQC be $n$ uncorrelated thermal qubits.
This state is related to $\rho^{BQC_P} = |00 \cdots 0\>\<00 \cdots 0|$
through an independent bit flip error operation ${\cal E}$ such that
$\rho^{BQC} = {\cal E}(\rho^{BQC_P})$, where
$ {\cal E}(\rho) \equiv (\otimes_{i=1}^n {\cal E}_i) (\rho) 
	= \sum_k p_k \sigma_x^k \rho \sigma_x^k 
$
with ${\cal E}_i (\rho) \equiv q_i \rho+(1- q_i)\sigma_{x_i} \rho
\sigma_{x_i}$.
Let the measurement signal from the $i^{th}$ qubit be
$E_i(\rho)=\<\sigma_{z_i}\> = tr(\rho \sigma_{z_i})$. 
A  quantum algorithm $U$ is an HQA if it satisfies the following two
conditions.
(1) BQC$_P$ can implement $U$ with at most polynomial slowdown compared with
    using the SQC model. 
(2) $E_i({\cal F}(\rho)) = c_iE_i(\rho)$ where $c_i$ are independent of
$n$ and ${\cal F}$ is defined by ${\cal U}\circ {\cal E} = {\cal
F}\circ {\cal U}$ with ${\cal U}(\rho) = U \rho U^{\dagger}$.
Note, when ${\cal U}$ and ${\cal E}$ are given,
${\cal F}$ is determined by ${\cal F} (\rho) \equiv \sum_k  B_k\rho
B_k^{\dagger}$ with $B_k = \sqrt{p_k} U \sigma_x^k U^{\dagger}$. 

We show here that condition (2) ensures BQC and BQC$_P$ are
polynomially equivalent.  
Since the initial state of the BQC is $\rho^{BQC} = {\cal
E}(\rho^{BQC_P})$, implemenenting ${\cal U}$ in BQC is equivalent to
implementing ${\cal U}\circ {\cal E}$ in BQC$_P$.
By condition (2), ${\cal U}(\rho^{BQC}) = {\cal F}\circ
{\cal U} (\rho^{BQC_P})$ and 
$
E_i( {\cal U} (\rho^{BQC})) = c_i E_i({\cal U}(\rho^{BQC_P}).
$
In other words, under condition (2), the initial state imperfection
only causes a constant signal loss without changing the output
information, hence, polynomially relating BQC and BQC$_P$. 
Together with condition (1), BQC and SQC are polynomially related in 
implementing HQA. 

We now consider what algorithms are HQAs using the criteria we have developed. 
From previous discussions, it follows that the DJ algorithm is an HQA with 
${\cal F} = {\cal E}$ and $E_i({\cal F}(\rho)) = (2q_i-1)E_i(\rho)$.
Another problem solved by an HQA is the parity problem \cite{Cleve97},
which determines an unknown $y\in \{0,1\}^n$ using the minimum number
of queries of the inner product function $f: \{0,1\}^n \rightarrow
\{0,1\}$, $f(x) = x \cdot y $.
This problem is solvable by the same procedure given by
Eqs.~(\ref{eq:step1})-(\ref{eq:step3}). 
Now with $f(x) = x\cdot y$, the final state following 
from Eq.~(\ref{eq:step3}) is,
\be	
	\frac{1}{2^n}\sum_{x,z \in \{0,1\}^n }(-1)^{x\cdot (y\oplus z)}
	\ket{z}\ket{0} = \ket{y}\ket{0}
\,.
\ee
BQC outputs $E_i^{BQC} = (2q_i-1) E_i^{BQC_P}= (2q_i-1) y_i$. 
Note that in contrast to the DJ problem, the parity problem has a
deterministic answer such that no classical amplification is needed.
Since the reduction in the signal, $2 q_i - 1$, is independent of $n$,
there is sufficient confidence to determine $y$ with only {\em one}
query, the {\em same} as for the SQC model.
Classical information theory, however, shows at least $n$ queries are
needed when $y$ is randomly distributed.

Condition (1) for an algorithm to be an HQA may seem to be overly
restrictive, but in reality it is not.  We have found efficient
modifications of all known quantum algorithms\cite{Cleve97,QAlist} for the
BQC$_P$ model.  
The general method for doing this, as pointed out originally in
\cite{Gershenfeld97}, is to perform the usual classical
post-computation (such as continued fraction expansions, and inner
products) as part of the quantum algorithm, on each quantum computer
in the ensemble.
Furthermore, since these quantum algorithms generally produce
solutions which are efficiently verifiable, this procedure in effect
{\em determinizes} the quantum algorithm and translates probabilistic
operation into a bounded constant signal loss.

Nevertheless, the importance of condition (1) for an algorithm to be
an HQA is paramount, due to the difficulty to satisfy both conditions 
(1) and (2) simultaneously. 
For example, Simon's algorithm~\cite{Simon94}, unmodified, satisfies
condition (2).
However, BQC$_P$ outputs a correct answer only for a few cases, but 
otherwise erases all useful information on the answer.  
Therefore, BQC$_P$ fails to solve Simon's problem with the original
algorithm even though BQC$_P$ and BQC are polynomially related.
Meanwhile, the modified Simon's algorithm for BQC$_P$ does not satisfy
condition (2).
Thus, we have not yet been able to construct an HQA for Simon's
problem.

HQA is easily generalized to computation models in which the
elementary carriers of quantum information are $q$-ary states for $q$
being a prime power.  In this extension,
the bit flip $\sigma_x$ is replaced by the generalized Pauli operator
$X_q$ defined as $X_q \ket{i} = \ket{i+1}$, where addition is taken
to be modulo $q$.  Then, ${\cal E}$ relating the BQC initial state  
$\rho^{BQC}$ to $|00\cdots 0\>$ can similarly be constructed.
When implementing the algorithm $U$, BQC and BQC$_P$ are polynomially
related provided the corresponding operation ${\cal F}$ satisfies 
$E_i({\cal F}(\rho)) =c_i E_i(\rho)$, with 
$c_i$ independent of $n$.  An algorithm satisfying
this is $\tilde{U} = (DFT_q^{-1})^{\otimes n} \tilde{U}_f
(DFT_q)^{\otimes n}$
where $DFT_q$ denotes the discrete Fourier Transform, 
$DFT_q \ket{x} = \frac{1}{\sqrt{q}} \sum_{y=0}^{q-1} e^{i2\pi x\cdot y
/q} \ket{y}$, and $\tilde{U}_f$ commutes with the phase 
operator $Z_q$, generalized similar to $X_q$.

In conclusion, we have shown the existence of a class of algorithms,
HQA, which are polynomially equivalently realizable with the standard
quantum computer model, and the bulk quantum computer model with
either pure or thermally mixed inputs.  These algorithms can thus
retain many advantages provided by quantum algorithms over the
classical, even when implemented on physical systems with limited
capabilities.  In particular, for non-trivial algorithms in HQA such
as the DJ algorithm, which require entangling unitary operations
(since single qubit operations can only implement $2(2^n-1)$ balanced
functions out of ${2^n \choose 2^{n-1}}$ in total), the necessity of
such nonclassical operations implies that classical simulation of BQC
would be difficult even though BQC might involves only separable
states \cite{Schack99}.
HQA can be generalized in several ways. 
First, the BQC initial state can be any state, related to $|00 \cdots 0\>$ 
through a general error operation ${\cal E}$, other than the thermal state. 
Second, modifications to the original algorithm and non-unitary
initialization can be performed as long as they do not detract from
the complexity of the original algorithm $U$.
More concretely, let $U$ satisfy condition (1).
If there exists an operation ${\cal V}$, possibly non-unitary,
such that ${\cal F}$, defined by 
${\cal V}\circ {\cal E} = {\cal F} \circ {\cal U}$,
gives 
$E_i({\cal F}(\rho)) = c_i E_i (\rho)$ with $c_i$ independent of $n$, 
then $U$ is an HQA. 
The only constraint is that the implementations of ${\cal V}$ and
${\cal U}$ have polynomially equivalent complexity.
These results indicate that a deep relationship exists between hot qubit
algorithms and algorithms which do not spread initial errors, or algorithms
involving quantum error correction.
It is likely that a good place to look for applications of the BQC model would
be in quantum algorithms which are robust against the initial uncorrelated bit
flip errors.
Further study of such relationships may lead to other interesting quantum
algorithms for bulk quantum computation with imperfect input states, and other
physical models of quantum computation which are realizable but limited by
practical considerations.
We would like to thank Dr.~Hoi-Fung~Chau for his suggestion to
generalize HQA to higher dimensions. We thank Lieven Vandersypen for
helpful discussions and Prof. James Harris for support. This work was
supported by the DARPA Ultra-scale Program under the NMRQC initiative,
contract DAAG55-97-1-0341.


 

\end{multicols}
\end{document}